 \documentclass[sigconf, 9pt, nonacm]{acmart}
 
\AtBeginDocument{%
  }

\setcopyright{acmlicensed}
\copyrightyear{2018}
\acmYear{2018}
\acmDOI{XXXXXXX.XXXXXXX}
\acmConference[Conference acronym 'XX]{Make sure to enter the correct
  conference title from your rights confirmation email}{June 03--05,
  2018}{Woodstock, NY}
\acmISBN{978-1-4503-XXXX-X/2018/06}

\usepackage{tikz}
\newcommand*\circled[1]{\scriptsize{\tikz[baseline=(char.base)]{\node[shape=circle,draw,inner sep=2pt] (char) {#1};}}\normalsize}
\usepackage{algorithm}
\usepackage{algorithmic}
\usepackage{graphicx}
\usepackage{subcaption}

\begin{document}

\title{Edge-Oriented Orchestration of Energy Services Using Graph-Driven Swarm Intelligence}




\author{Liana Toderean}
\email{liana.toderean@cs.utcluj.ro}
\affiliation{
  \institution{Technical University of Cluj-Napoca}
  \country{Romania}
}

\author{Dragoș Lazea}
\email{dragos.lazea@cs.utcluj.ro}
\affiliation{
  \institution{Technical University of Cluj-Napoca}
  \country{Romania}
}

\author{Vasile Ofrim}
\email{vasile.ofrim@cs.utcluj.ro}
\affiliation{
  \institution{Technical University of Cluj-Napoca}
  \country{Romania}
}

\author{Stefania Dumbrava}
\email{stefania.dumbrava@ensiie.fr}
\affiliation{%
  \institution{ENSIIE \& SAMOVAR}
  \country{France}
}

\author{Anca Hangan}
\email{anca.hangan@cs.utcluj.ro}
\affiliation{%
  \institution{Technical University of Cluj-Napoca}
  \country{Romania}
}

\author{Tudor Cioara}
\email{tudor.cioara@cs.utcluj.ro}
\affiliation{%
  \institution{Technical University of Cluj-Napoca}
  \country{Romania}
}


\settopmatter{authorsperrow=3}

\renewcommand{\shortauthors}{Toderean et al.}

\begin{abstract}
As smart grids increasingly depend on IoT devices and distributed energy management, they require decentralized, low-latency orchestration of energy services. We address this with a unified framework for edge–fog–cloud infrastructures tailored to smart energy systems. It features a graph-based data model that captures infrastructure and workload, enabling efficient topology exploration and task placement. Leveraging this model, a swarm-based heuristic algorithm handles task offloading in a resource-aware, latency-sensitive manner. Our framework ensures data interoperability via energy data space compliance and guarantees traceability using blockchain-based workload notarization. We validate our approach with a real-world KubeEdge deployment, demonstrating zero-downtime service migration under dynamic workloads while maintaining service continuity.
\end{abstract}




\maketitle

\section{Introduction}

The Internet of Things (IoT) is transforming smart grids and energy services by enabling real-time monitoring and decentralized energy management using both traditional control and experimental ML approaches. However, these services face challenges due to the continuously growing volume of data generated by IoT devices and transmitted across the network to centralized cloud infrastructures. The requirements for low latency, real-time responsiveness, and localized decision-making in energy systems call for new architectures that can process data closer to the source, minimizing network congestion and improving operational efficiency~\cite{ullah2023optimizing}.


In this context, modern edge-fog-cloud infrastructures allow efficient distributed data processing along the entire computing continuum, spanning from resource-constrained edge devices to powerful centralized cloud data centers. Thus, identifying the optimal placement of tasks across available computing nodes is essential to reduce latency, balance computational load, and maximize resource efficiency across all layers. This becomes even more critical as resource-intensive tasks, such as Tiny-ML inference, become increasingly common on edge devices, requiring careful consideration of resource constraints, network conditions, and workload characteristics. Furthermore, in smart grid, critical services are safeguarding the operation of the infrastructure, therefore, their execution is not only constrained by the fast data availability from the meter but also by energy grid constraints and stringent time execution constraints. For example, congestion management services AI models forecast the available generation and consumption in a local microgrid; therefore, in their orchestration at the edge, we must go beyond the computational and data network factors and consider additionally electricity network-specific factors (e.g., real-time production vs. consumption balance, active vs reactive power, etc.) ~\cite{antonesi2025hybrid}. In case of congestion, without quick balancing, the grid can experience dangerous fluctuations and, in extreme cases, power outages. However, to our knowledge, nowadays orchestration platforms are not able to consider energy grid-specific factors in energy services offloading at the edge decision-making, representing a significant gap for the roll-up of AI-driven energy services in the smart grid.
Furthermore, Federated AI services represent another important category of energy-related services, enabling collaborative and privacy-preserving training of machine learning models across the smart grid~\cite{toderean2025heuristic}. In such cases, executing the task on a computing node located far from either the data source or the target destination can introduce higher communication latency and negatively impact the responsiveness and overall efficiency of grid operations~\cite{sahni2018data}. Although federated AI services in energy systems often consider computational topology, they typically overlook the physical structure of the grid, an underexplored but critical factor for informed decision-making.


To address the above-identified gaps, several research questions need to be addressed.


\circled{1} \textit{How can energy service tasks be efficiently offloaded to edge computing nodes in a way that meets resource constraints, minimizes communication latency, and preserves data locality, while additionally considering constraints related to grid topology, critical energy services requirements, and energy locality?} Finding an efficient placement of tasks on distributed computing nodes in edge-cloud infrastructures is a challenging problem that has been extensively studied from various perspectives in the literature~\cite{alameddine2019dynamic, chai2023joint, lee2021deadline, sahni2018data}. For example,~\cite{alameddine2019dynamic} jointly address offloading, resource allocation, and scheduling to meet IoT deadlines in collaborative edge environments. In\cite{chai2023joint}, the authors propose a multi-task offloading and resource allocation scheme that accounts for dependencies and resource constraints in a satellite IoT use case. The work in~\cite{lee2021deadline} introduces a heuristic scheduler focused on reducing the deadline miss ratio (DMR) by prioritizing deadlines and network flows. A proximity-based placement method is proposed in~\cite{sahni2018data}, using a multistage greedy algorithm to assign tasks near their data sources. Heuristic methods are increasingly favored due to the impracticality of exhaustive search in large systems~\cite{DBLP:journals/diot/LatipAHKG24}, though even these may suffer from high computational overhead when handling large numbers of tasks and nodes~\cite{DBLP:journals/dcan/MahengeLS22}. In this regard, optimization algorithms capable of handling multi-objective formulations and complex constraint definitions are essential for effective and adaptive task placement~\cite{liu2020leveraging}.


\circled{2} \textit{What data model can effectively support efficient topology exploration and offloading decision making?} Relational databases often face performance limitations when modeling topology and connectivity-centric data. In contrast, graph databases offer scalable and flexible querying for highly interconnected data~\cite{besta2023demystifying, de2015towards}. Several works have explored modeling network topologies using graph databases~\cite{de2015towards, bannour2022gox, jamkhedkar2018graph}. In~\cite{de2015towards}, the authors integrate the Network Markup Language (NML) into a graph database to enhance SDN network state and demonstrate efficient path queries. GOX~\cite{bannour2022gox} extends this by modeling SDN topologies with Neo4j, achieving improved performance on synthetic and real-world networks. Similarly, the work in~\cite{jamkhedkar2018graph} presents a graph-based system for managing topology and inventory in dynamic cloud networks to support automated management tasks. 


\circled{3} \textit{How can data interoperability and movement notarization for energy services be ensured to enable traceable and efficient offloading toward the edge?}  
Data interoperability is an essential requirement for energy services as they operate with heterogeneous data sources and they must understand and exchange data with other services ~\cite{molokomme2022edge}. Standardized data exchanges~\cite{minh2022edge, arcas2024edge} facilitated by data spaces frameworks provide interoperability for energy services through semantic models, identity management, and policies~\cite{coppolino2024exploiting}. Data space connector acts as a trusted interface for participants to exchange data in a secure and controlled manner. Blockchain technology provides tamper-proof tracking mechanisms that enable transparency and traceability ~\cite{10360267}. It serves as a platform for notarizing task offloading and monitoring resource usage in a secure and verifiable manner~\cite{9933642}. However, to maintain interoperability during service migration, data space connectors must be actively employed between services to preserve secure and policy-compliant connections. Additionally, by tokenizing tasks and leveraging smart contracts, the placement and execution of offloaded services should be automatically traced whenever a task is relocated, ensuring accountability and trust.

Motivated by the above, we explore how to efficiently offload energy service-related tasks to computing nodes in a heterogeneous edge–fog–cloud system integrated with the smart energy grid infrastructure. A graph-based data model is defined to represent both computational infrastructure and workload tasks, facilitating the network topology exploration. It supports the execution of heuristic optimization algorithms by effectively defining and constraining the computational search space and ensuring that the algorithm operates only on data relevant for generating feasible offloading solutions. 
It enables orchestration with computational interoperability, while data space compliance and blockchain tracing ensure data interoperability.

This work presents three key contributions: (a) the design of an edge-oriented orchestration framework for delivering energy services to edge nodes in the smart grid, while ensuring service continuity, data interoperability, and availability across distributed resources; (b) a graph-based model that unifies the physical infrastructure, such as computing nodes and IoT smart meters, and the workload consisting of energy services' tasks, enabling efficient graph querying for network exploration and data reduction; and (c) an edge-aware, heuristic-based optimization technique that efficiently places energy-related tasks across heterogeneous edge computing nodes, optimizing for latency, energy consumption, and resource constraints to improve service delivery.
\begin{figure}[t]
    \centering
    \includegraphics[width=0.8\linewidth, trim=6cm 14cm 8cm 3cm, clip]{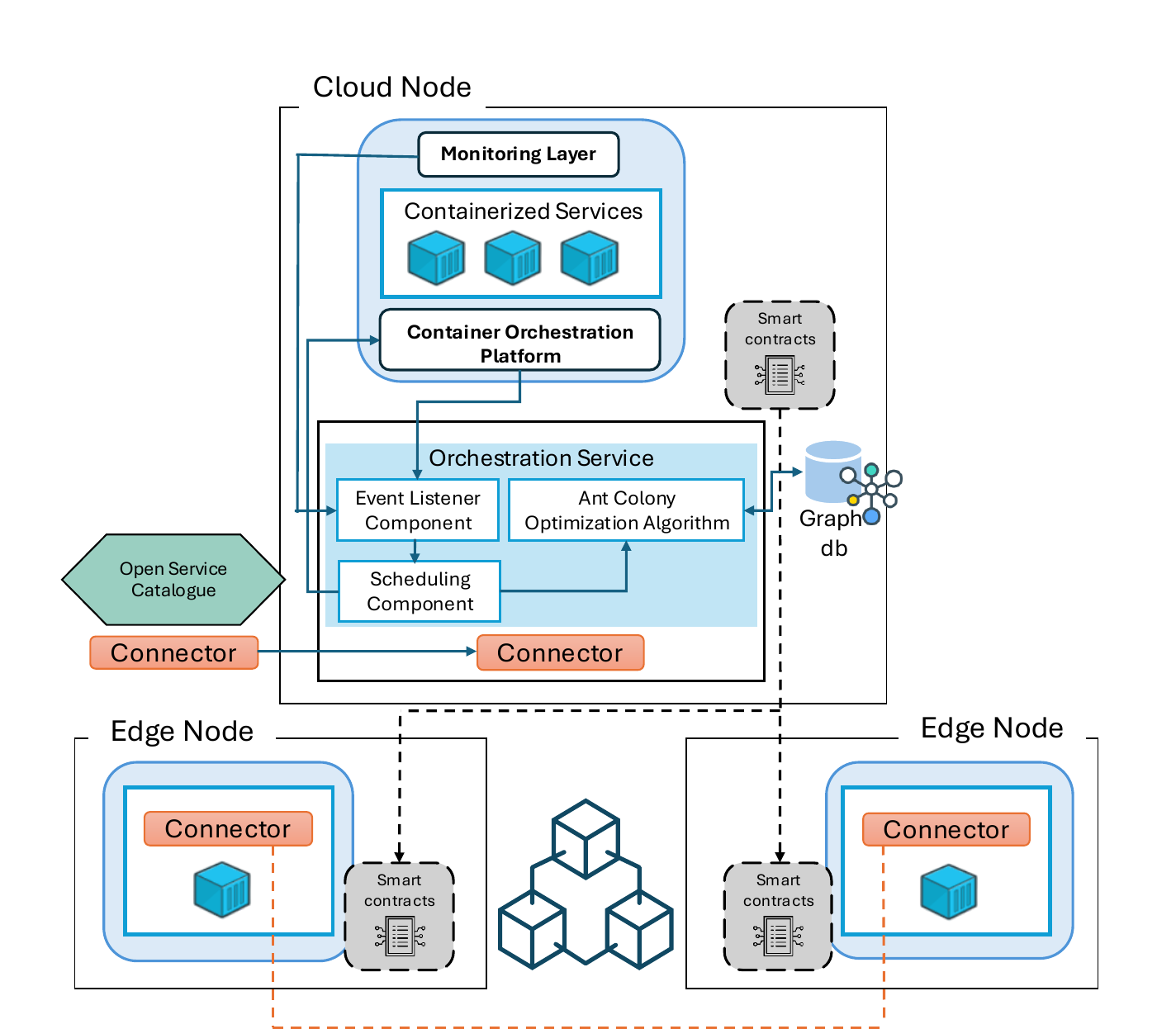}
    \setlength{\belowcaptionskip}{-20pt}
    \caption{Orchestration framework architecture}
    \label{fig:orch}
\end{figure}

\section{Orchestration and Data Management}

This section outlines the orchestration framework and data management mechanisms that support efficient orchestration and trustworthy execution of energy services. A unified graph-based model integrates infrastructure and workload information to enable scalable task placement via in-graph filtering and swarm-based optimization. Data interoperability is ensured using data space standards, while blockchain-based tracing guarantees transparency and auditable task execution across the edge–fog–cloud continuum.

The orchestration framework architecture is depicted in Figure 
\ref{fig:orch}. It is designed to support the deployment and management of energy services across a dynamic, distributed cloud-edge continuum whilst ensuring service availability and preserving the connection among them. The connection between services is managed through data space connectors, and the services' metadata is exposed through a federated catalogue that enables the orchestration service to access it. The framework enhances the underlying container orchestration platform by introducing a monitoring layer that observes the status of nodes and listens for specific events regarding the computational resources on nodes or network conditions. The scheduling component triggers the swarm-based optimization algorithm that generates a new task scheduling plan that is aligned with the application requirements in terms of computational resources, network conditions, and geographical considerations. The container orchestration platform handles the  actual service migration, according to the generated schedule with the new state of the system being updated in the graph database. It continues to monitor, collect and aggregate metrics from nodes and their workload, providing real-time insights to the orchestration service.



\subsection{Graph Data Model and Task Placement}\label{ssec:graph_model}

\begin{figure}[t]
    \centering
    \includegraphics[width=0.65\linewidth]{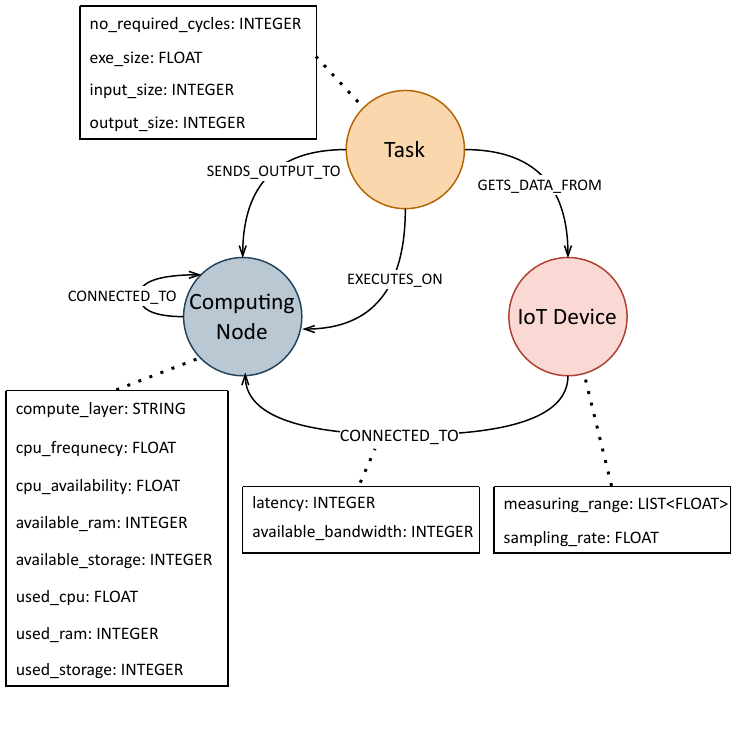}

    \caption{Unified graph model}
    \label{fig:datamodel}
\end{figure}

\textbf{Graph Data Model.} We define a graph data model that provides a unified representation of both the infrastructure and the workload, as depicted in Figure~\ref{fig:datamodel}. The model captures details about computational resources—both available and in use—on each node, along with the connectivity between nodes. Links include attributes such as latency and available bandwidth, enabling a comprehensive view of the system’s state. The workload is associated with the infrastructure by defining data dependencies—specifically, the IoT device (e.g., a smart meter) that generates the input data required by the task, and the computing node where the task’s output is intended to be delivered. Furthermore, tasks assigned to execution nodes as a result of the heuristic-based optimization are explicitly linked to those nodes within the model by creating \verb|EXECUTES_ON| relations. This relation might not always be present, as some tasks may remain unassigned.

This model enables efficient topology exploration and path querying when identifying suitable execution nodes for task placement, while also supporting preliminary data filtering before applying the optimization algorithm. This is particularly important in large-scale infrastructures where loading the entire topology in the optimization engine is neither practical, nor efficient. To this aim, we implement a filtering mechanism that, for each unassigned task, reduces the dataset by loading into the optimization module only those nodes that: \circled{1} have a path to the IoT device generating the task’s input data, \circled{2} are connected—either directly or indirectly—to the node where the task’s output should be delivered, and \circled{3} have enough available resources to accommodate the task. Latency and average bandwidth for input/output paths are computed via graph queries and stored with each node in the filtered set. This allows the optimization engine to select nodes from a pre-filtered, feasible subset, reducing search space and computational cost.


\textbf{Formal Problem Definition.}  We formulate the task-to-node mapping as a \textit{multi-objective optimization problem}. Let the edge-cloud infrastructure be \( I = D \cup E \cup F \cup C \), where \( D = \{d_1, d_2, \ldots, d_{N_d}\} \) denotes the set of IoT devices (e.g., smart meters), \( E = \{e_1, e_2, \ldots, e_{N_e}\} \) is the set of edge computing nodes, \( F = \{f_1, f_2, \ldots, f_{N_f}\} \) is the set of fog computing nodes, and \( C = \{c_1, c_2, \ldots, c_{N_c}\} \) is the set of cloud computing nodes. Given a set of tasks \( T = \{t_1, t_2, \ldots, t_{N_t}\} \), the objective is to assign each task \( t_i \in T \) to a computing node \( n_j \in E \cup F \cup C \) such that multiple objectives—minimizing execution time, communication latency, and energy consumption—are optimized while satisfying the resource and connectivity constraints of the infrastructure.

Each task \( t_i \in T \) is characterized by a set of parameters: \( C_i \), representing the required CPU cycles; \( S_i^{in} \), the input data size; \( S_i^{out} \), the output data size; and \( S_i^{exe} \), the executable size. Similarly, each computing node \( n_j \in E \cup F \cup C \) is defined by its CPU frequency, $freq_j$, and its available resources (i.e., the difference between the total amount of resources at the node and the amount of resources already used), including \( R_j^{CPU} \) for CPU availability, \( R_j^{RAM}  \) for available RAM memory, and \( R_j^{storage} \) for available storage. The available resources at each node are updated after assigning a task by subtracting the resources consumed by that task from the current available resources of the node. In addition to its available and in-use resources, each node $n_j$ is also characterized by a coefficient $k_j$, which reflects the CPU architecture—dependent energy consumption. This coefficient is used to compute the energy consumed during the execution of tasks on $n_j$.
Each task--candidate node pair \( (t_i, n_j) \) is associated with two communication paths: \( P_{ij}^{in} \), the path from the data source of task \( t_i \) to the candidate execution node \( n_j \), and \( P_{ij}^{out} \), the path from \( n_j \) to the node where the output of \( t_i \) must be delivered. Each path has an associated total latency and average available bandwidth: \( L_{ij}^{in} \) and \( BW_{ij}^{in} \) for the input path, and \( L_{ij}^{out} \) and \( BW_{ij}^{out} \) for the output path. These attributes are computed during the in-database filtering step performed.

The goal is to compute a task-to-node mapping that minimizes the average cost across all assignments, where the cost $cost_{ij}$  of mapping task $t_i$ to node $n_j$ integrates execution time, communication delay, and energy consumption. This optimization considers only valid assignments, i.e., those that respect node resource constraints, while aiming to improve overall system efficiency.
{\setlength{\abovedisplayskip}{0pt}
 \setlength{\belowdisplayskip}{0pt}
\begin{align}
\text{exec}_{ij} &= \frac{C_i}{freq_j} \label{eq:exec} \\
comm_{ij} &= \frac{S^{in}_i}{BW^{in}_{ij}} + L^{in}_{ij} + \frac{S^{out}_i}{BW^{out}_{ij}} + L^{out}_{ij} \label{eq:comm} \\
\text{energy}_{ij} &= k_j \cdot C_i \cdot (freq_j)^2 \label{eq:energy} \\
\text{cost}_{ij} &= w_{ex} \cdot \text{exec}_{ij} + w_{co} \cdot \text{comm}_{ij} + w_{en} \cdot \text{energy}_{ij} \label{eq:cost}
\end{align}}

Here, $\text{exec}_{ij}$ and $\text{energy}_{ij}$ are, respectively, the execution time and energy consumed when running task $t_i$ on node $n_j$, and $\text{comm}_{ij}$ is the communication time along \( P_{ij}^{in} \) and \( P_{ij}^{out} \). 

\begin{algorithm}[b]
\caption{ACO for Task-to-Node Assignment}
\begin{algorithmic}[1]
\label{aco-alg}
\REQUIRE 
Tasks $T$, candidate node map $O$, cost matrix $cost_{ij}$, ACO params $(\alpha, \beta, \rho, \tau_0)$, grid topology $G$
\ENSURE Mapping $M: T \rightarrow N$ minimizing average cost

\STATE Initialize pheromone matrix $\tau_{ij} \leftarrow \tau_0$ for all valid $(t_i, n_j)$
\STATE Compute heuristic $\eta_{ij} \gets \eta(BW_{ij}, L_{ij}, G)$

\FOR{each iteration}
  \FOR{each ant}
    \STATE Initialize $S \gets \emptyset$
    \FOR{each task $t_i \in T$}
      \STATE Compute probabilities: $p_{ij} = \frac{\tau_{ij}^\alpha \cdot \eta_{ij}^\beta}{\sum_{k \in \text{valid}} \tau_{ik}^\alpha \cdot \eta_{ik}^\beta}$
      \STATE Select node $n_j$ from $O(t_i)$ based on $p_{ij}$
      \STATE Update $R_j^{CPU}$, $R_j^{RAM}$, $R_j^{storage}$, $S \gets S \cup \{(t_i, n_j)\}$
    \ENDFOR
    \STATE Compute total cost of $S$: $\frac{1}{|T|} \sum_{(t_i,n_j) \in S} cost_{ij}$
    \STATE Update best solution if needed
  \ENDFOR
  \STATE Evaporate pheromones: $\tau_{ij} \leftarrow (1 - \rho) \cdot \tau_{ij}$
  \STATE Reinforce pheromones using best solution
\ENDFOR
\RETURN $M$
\end{algorithmic}
\end{algorithm}

\textbf{Swarm-based Optimization for Task Offloading.} We address the task-to-node assignment problem using a custom Ant Colony Optimization (ACO) algorithm~\cite{dorigo2007ant}, due to its ability to effectively handle discrete problems, incorporate custom heuristics, and maintain a dynamic balance between exploration and exploitation through pheromone updates. Unlike Genetic Algorithms~\cite{forrest1996genetic}, which may require complex encoding of problem variables and risk premature convergence, or Simulated Annealing~\cite{bertsimas1993simulated}, which converges slowly raising scalability issues and struggles with defining constraints, ACO efficiently integrates constraints and adapts to dynamic environments with minimal overhead. Its scalable design and memory-like pheromone mechanism make it a suitable choice for dynamic, resource-constrained mapping scenarios, such as the assigning computing tasks to executing nodes problem.

As part of our approach, each artificial ant builds a solution by probabilistically selecting execution nodes for tasks, guided by pheromone trails and a heuristic favoring high bandwidth and low latency paths: $\frac{BW_{ij}^{in} + BW_{ij}^{out}}{L_{ij}^{in} + L_{ij}^{out}}$. Only resource-respecting mappings are considered. Pheromone values are updated after each iteration to reinforce good solutions while enabling exploration. Over multiple iterations, the algorithm converges to a mapping that minimizes execution, communication, and energy costs. Algorithm~\ref{aco-alg} summarizes our approach.

\begin{figure}[t]
    \centering
    \includegraphics[width=0.8\linewidth]{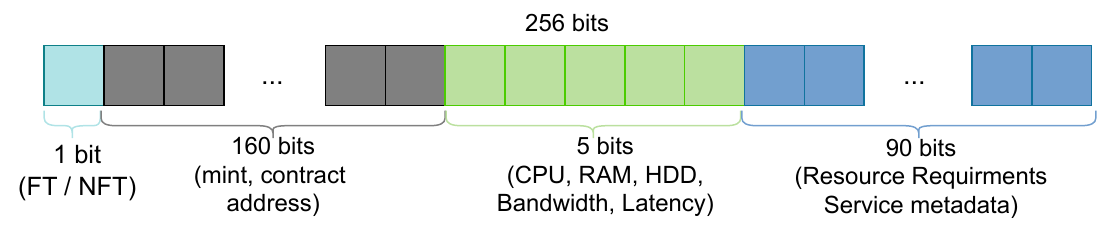}   
    \setlength{\belowcaptionskip}{-20pt}
    \caption{Structure of a Task Token}
    \label{fig:token}
\end{figure}

\subsection{Data Interoperability and Task Tracing}\label{ssec:data_interop}

Data interoperability is essential for digital energy services, enabling the integration of diverse systems and devices. Effective coordination of multiple stakeholders (grid operators, market platforms, regulatory entities) requires data exchange between services, IoT devices, and external forecasting systems. Processing data from these heterogeneous systems is important for real-time decision-making. Data must be shared reliably and with controlled access. Beyond data interoperability, edge deployment of energy services requires a decentralized mechanism to ensure transparency and traceability of task location and duration.

Data space standards (defined by IDSA) offer a framework for secure data exchanges through data connectors that enable standardized, policy-enforced communication. Also, they support integration with a federated service catalog enabling easier discovery. Some of the requirements of energy services include reliable, secure communication and interoperability. The connector is used for communication between services as well as by the orchestrator to discover metadata exposed through the service catalog, information that guides the orchestration process. The energy services often act as both consumers and producers of information for other services. 
Services that will share data through the connector define an asset that describes the data it intends to share. This asset includes metadata (i.e., description of the exposed data), access methods (i.e., push- or pull-based), usage, and policy constraints. By defining the assets and access policies, the data producers have a standardized way to share their data whilst maintaining control over it and ensuring policy enforcement. When another service seeks access to this data, it first discovers the asset using the federated catalog and then initiates a contract negotiation process. This negotiation part ensures that both parties are authorized and all policies and enforcements are respected and then a contract agreement will be generated that manages the transfer. This way, data consumers can have controlled access according to their usage rights to data from multiple providers that it is ensured to be data space compliant. 

The task representation on the blockchain platform is made through a multi-type token that enables efficient single and batch transfers. The token id structure can be used to incorporate metadata about the task, making it traceable and verifiable on chain. An example of the id structure is presented in Figure \ref{fig:token} where the first bit is used to indicate the token type. The fungible tokens (FT) are used to automate payments for the period that the task was running on a resource, whilst the non-fungible tokens (NFT) are the ones representing the tasks. For NFTs, the remaining bits encode the mint address (i.e., the requesting entity or contract), required resources, resource limits (for selected types), and task metadata such as location, linked services, or ownership.

\begin{figure}[h!]
    \centering
    \includegraphics[width=0.7\linewidth]{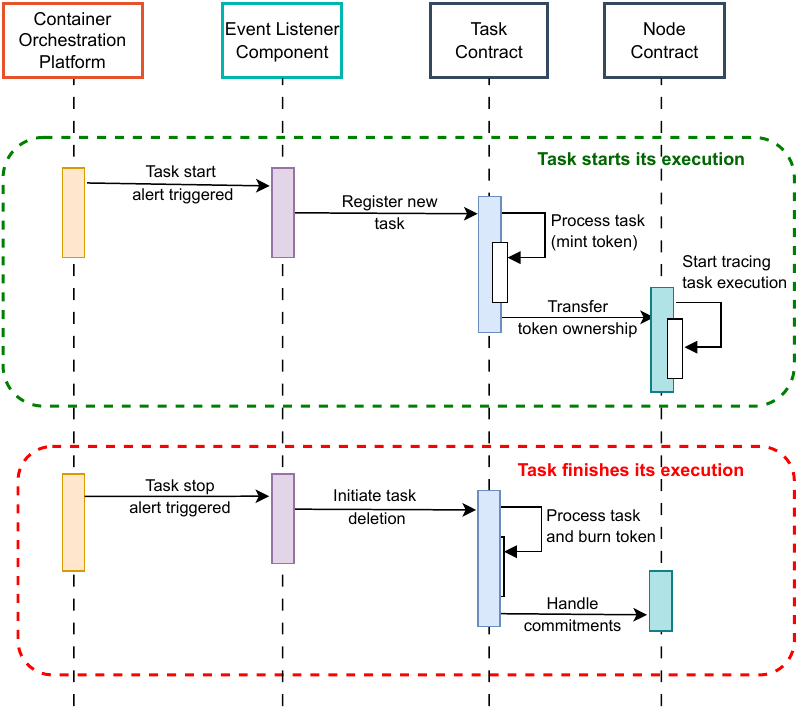}
   \setlength{\belowcaptionskip}{-10pt}
    \caption{Blockchain task tracking flow}
     
    \label{fig:blockhain-flow}
\end{figure}

Smart contracts manage the interaction of edge nodes and tasks on chain. Whenever a service is deployed, a token is minted to represent that running task. The task smart contract ensures that the tokens have a valid structure and only authorized parties are able to mint them. The token is sent to the smart contract that represents the edge node where the service is deployed. The node tracks resource usage by the task using information stored in the received NFT token. Based on that, it can enforce automatic payment for resource utilization (depending on the usage policies) before releasing the token. This prevents the task from being relocated before the commitments are fulfilled. The token is transferred to another edge node when the service is relocated and is destroyed when the service stops, as shown in Figure \ref{fig:blockhain-flow}. This ensures traceability and enables the service and infrastructure providers to track, monitor, and enforce automated payment for the resource usage. 

\section{Preliminary Solution Validation}\label{sec:experiments}


\begin{table}[b]
\centering
\small

\caption{Testbed Infrastructure Configuration}
\begin{tabular*}{0.48\textwidth}{@{\extracolsep{\fill}}lcc}
\toprule
\textbf{Node Type} & \textbf{CPU} & \textbf{RAM} \\
\midrule
Cloud Node    & i5-7400 (4 cores, 3.0 GHz) & 16 GB \\
Edge Node 1   & i3-7100U (2 cores, 2.4 GHz) & 8 GB \\
Edge Node 2   & i5-5200U (2 cores, 2.2 GHz) & 6 GB \\
Edge Node 3   & i5-5200U (2 cores, 2.2 GHz) & 8 GB \\
\bottomrule
\end{tabular*}

\label{tab:gas_cosumption}
\end{table}

We validate our solution through a complete orchestration flow, starting from a Neo4j model of the edge-cloud infrastructure, which feeds an ACO engine for efficient task-to-node mapping and optional remapping. Placement decisions update the graph and are applied on a KubeEdge testbed in a distributed energy scenario, demonstrating effective handling of dynamic workloads (Table~\ref{tab:infrastructure}). To assess scalability beyond the testbed’s  scale and heterogeneity, we generate synthetic infrastructures with varied node types and task profiles, validating both the optimizer and its integration with Neo4j. Next, we demonstrate the framework's ability to manage dynamic workloads on physical infrastructure.



\begin{figure*}[t]
    \centering
    \begin{subfigure}[t]{0.32\textwidth}
        \includegraphics[width=\linewidth]{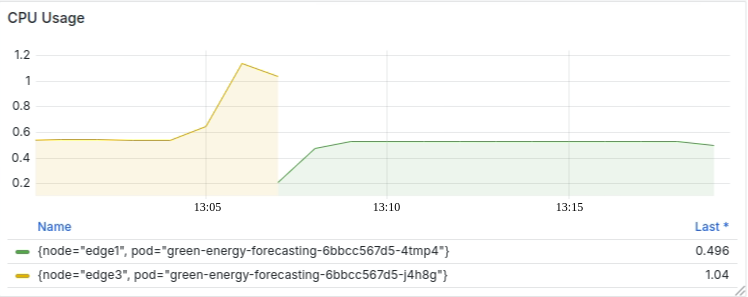}
        \caption{CPU usage before/after migration}
        \label{fig:pod_migration}
    \end{subfigure}
    \hfill
    \begin{subfigure}[t]{0.32\textwidth}
        \includegraphics[width=\linewidth]{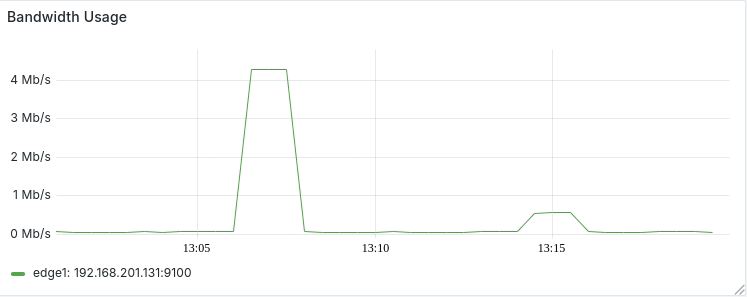}
        \caption{Migration Bandwidth Overhead}
        \label{fig:pod_migration_overhead}
    \end{subfigure}
    \hfill
    \begin{subfigure}[t]{0.32\textwidth}
        \includegraphics[width=\linewidth]{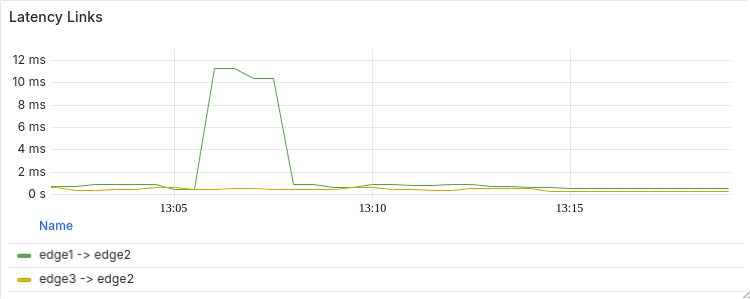}
        \caption{Pod Migration Latency Overhead}
        \label{fig:pod_migration_latency_overhead}
    \end{subfigure}
      \setlength{\belowcaptionskip}{-10pt}
    \caption{Overview of pod migration performance: CPU, bandwidth, and latency overhead}
    \label{fig:pod_migration_combined}
\end{figure*}


A machine learning-based energy management application is deployed via the Open Service Catalogue, comprising four subservices: (1) Energy Balancer, which coordinates energy consumption and production by processing real-time and forecasted data; (2) Photovoltaic Manager, which supplies real-time solar data to the Balancer; (3) Green Energy Forecasting, which uses weather data to predict solar output; and (4) Load Forecasting, that predicts energy consumption based on historical data. They are initially distributed across the infrastructure based on resource availability and proximity to data sources. Each is deployed along with its own Eclipse Dataspace Connector (EDC) to facilitate communication, adopting pull-based transfers for data requests (the Energy Balancer pulls from the Photovoltaic Manager) and push-based transfers for updates (the Green Energy and Load Forecasting services push to the Energy Balancer). EDC enforces a contract-based data exchange mechanism, where data sharing is allowed only between authorized participants under usage policies. These contracts persist beyond service or connector migrations, thus ensuring the continuity of data exchange regardless of deployment changes.

\begin{table}[h!]
\centering
\small
\caption{Gas consumption for token transactions}
\begin{tabular*}{0.48\textwidth}{@{\extracolsep{\fill}}cc}
\toprule
\textbf{Transaction Types} & \textbf{Gas consumption (gas units)} \\
\midrule
Mint    & 144373 \\
Transfer   & 56072 \\
Burn   & 29175 \\
\bottomrule
\end{tabular*}

\label{tab:infrastructure}
\end{table}

\vspace{-10pt}
For the Load Forecasting subservice we considered the transformer machine learning model described in ~\cite{antonesi2025hybrid}. The model is trained on historical energy consumption data, measured in kilowatt-hours (kWh), aggregated at hourly intervals. For each prediction, it uses the hourly energy consumption from the previous 7 days (168 values) as input and forecasts the hourly energy consumption for the next day (24 values). A data collector service gathers data from energy meters, performs the aggregation and stores the hourly energy values in a database. Considering this, the Load Forecasting service placement must be close to the data source. Also, the output of this services is used by the Energy Balancer service. The Load Forecasting service is scheduled daily to predict the energy consumption for the next day. The evaluation of resource utilization for this type of machine learning service is presented in Table ~\ref{tab:load_forecasting_resource_consumption}, showing moderate CPU and memory usage, a reasonable processing duration with minimal impact on latency between nodes and bandwidth consumption.

\begin{table}[h!]
\caption{Load Forecasting Service Requirements}
\centering
\begin{tabular*}{0.48\textwidth}{@{\extracolsep{\fill}}lllll}
\toprule
\textbf{Duration} & \textbf{CPU} & \textbf{RAM} & \textbf{Latency}  & \textbf{Bandwidth} \\
\midrule
1236 ms & 37.8\% & 20.5 MB & 200 ms & 20 KB/s \\
\bottomrule
\end{tabular*}
\label{tab:load_forecasting_resource_consumption}
\end{table}

We overloaded the Green Energy Forecasting subservice on Edge Node 3 with complex data streams. As the subservice's CPU demand exceeded the processing capacity of the node, a CPU pressure alert was triggered and forwarded by the Alert Manager. The alert was received by the Orchestrator, which planned and executed a redistribution, moving Green Energy Forecasting to Edge Node 1 using a scaling-based strategy for execution that maintained the service's high availability with zero downtime.
The migration process introduces brief bandwidth and latency overhead due to the subservice's host image pulling. As shown in Figure~\ref{fig:pod_migration_overhead}, bandwidth peaks at 4.0 Mb/s for a minute, without disrupting other services or causing network congestion. Similarly, Figure~\ref{fig:pod_migration_latency_overhead} shows a temporary inter-node latency spike up to 10 ms, with negligible impact. Figure~\ref{fig:pod_migration} shows CPU usage before and after migration. The Green Energy Forecasting service runs smoothly without resource starvation, confirming the framework's ability to handle resource pressure alerts and maintain continuous operation with minimal overhead. The Alert Manager emits events when a subservice changes its status. During migration, these trigger ownership transfer of the token minted at subservice start, enabling accurate tracking of the Green Energy Forecasting subservice’s location.
When the subservice  stops, its token is burned. The gas consumption for token operations on-chain is given in Table~\ref{tab:gas_cosumption}.


\textbf{Discussion}. The proposed orchestration framework can be extended for federated learning services to coordinate and ensure efficient decentralized machine learning training across edge nodes in the smart grid. The orchestrator service requires information about the computational resources available and statistical energy data from the edge nodes. In addition, it gathers information about the federated process regarding the hyperparameters used for model training and their distribution. Then, the orchestrator performs clustering and hyperparameter tuning to improve the communication and computational efficiency. For example, in a peer-to-peer federated learning process the orchestrator can increase communication efficiency by clustering edge nodes based on their characteristics. The edge nodes with similar profiles are grouped together and they communicate only with their clustering neighbors reducing communication overhead whilst simultaneously addressing the non-IID characteristic of energy timeseries. In addition, by performing hyperparameter tuning, it can improve the federated process by increasing the convergence speed and balance the local training across clients with different characteristics or energy profiles.
\section{Conclusions}\label{sec:conclusion}

We presented a graph-driven orchestration framework designed to support energy services across edge–fog–cloud infrastructures, combining efficient task placement via swarm intelligence with interoperable, traceable execution. Real-world validation on a KubeEdge testbed confirms its effectiveness under dynamic load with zero downtime. Key challenges ahead include scaling to larger topologies, adapting to highly volatile workloads, and automating policy-compliant service migration across heterogeneous administrative domains.

\small
\section*{Acknowledgments}
This research received funding from the European Union’s Horizon Europe research and innovation program under the Grant Agreements number 101136216 (Hedge-IoT). Views and opinions expressed are, however, those of the author(s) only and do not necessarily reflect those of the European Union or the European Climate, Infrastructure, and Environment Executive Agency. Neither the European Union nor the granting authority can be held responsible for them.

\bibliographystyle{ACM-Reference-Format}
\bibliography{bibfile}

\end{document}